# Single gap s-wave superconductivity in $Nb_2PdS_5$

Shruti[1], R. Goyal[2], V. P. S Awana[2], and S. Patnaik[1]

[1]School of Physical Sciences, Jawaharlal Nehru University, New Delhi 110067, India

[2]CSIR-National Physical Laboratory, Dr. K. S. Krishnan Marg, New Delhi 110012, India

**Abstract**

Superconducting order parameter and its symmetry are important parameters towards deciphering the pairing mechanism in newly discovered superconducting systems. We report a study on penetration depth measurement on $Nb_2PdS_5$ that has recently been reported with extremely high upper critical field with possible triplet pairing mechanism. Our data show that at low temperatures the change in penetration depth $\Delta\lambda$ is best fitted with BCS s-wave model for single gap. The zero-temperature value of the superconducting energy gap is estimated to be $\Delta_0$ = 1.05 meV. The superfluid density in the entire temperature range is well described by single gap with gap ratio $2\Delta_0/k_B T_c$ = 4 for $\lambda(0)$ = 220 nm.

**Introduction**

Recently superconductivity was reported in quasi one-dimensional compound $Nb_2PdS_5$ with superconducting transition temperature ($T_c$) of ~ 6 K and extremely high upper critical field ($H_{c2}$) >35 T [1]. This value exceeds the Pauli paramagnetic limit $H_{c2}(0) = 1.84 T_c$ by over 3 times [1-4]. The anisotropic upper critical field ($\gamma$ ~ 3) along with temperature dependent anisotropy was indicative of possible multiband superconductivity [1,5]. Concepts related to strong spin-orbit scattering and spin triplet pairing mechanism were invoked to explain the highest ever reported $H_{c2}(0)/T_c$ value [4,5]. The high $H_{c2}$ value and the presence of chalcogen in the layered structure are reminiscent of Chevrel phase compounds such as $PbMo_6S_8$ [6,7] and $SnMo_6S_8$ [8] but still there is no clarity on the macroscopic origin of superconductivity in $Nb_2PdS_5$.

The band structure calculations show that the Fermi surface of $Nb_2PdS_5$ is composed of multiple sheets with both quasi-two-dimensional (Q2D) hole character and quasi-one-dimensional (Q1D) sheets of both hole and electron character that may give rise to multiband superconductivity [1,5]. Multiband superconductivity is also predicted for similar quasi one-dimensional isomorphic $Nb_2Pd_xSe_5$ and $Ta_2Pd_xS_5$ compounds [9,10]. These aspects and concomitant symmetry of superconducting order parameter in $Nb_2PdS_5$ are still open questions. As evidenced in many superconducting systems, radio frequency (RF) penetration depth study has been proved to be extremely useful in establishing d wave superconductivity in cuprates [11,12], s± wave symmetry in iron pnictide [13] and two gap superconductivity in $MgB_2$ superconductors [14]. In this paper we report a study on the symmetry of the superconducting order parameter and superconducting gap by penetration depth measurement of $Nb_2PdS_5$ using tunnel diode oscillator technique. Results from our study show that the data are best described by s-wave model with superconducting gap $2\Delta_0/k_B T_c = 3.9$ which is higher than the weak coupling BCS value. This suggests that $Nb_2PdS_5$ to be an intermediate-strong coupling superconductor.

**Experimental**

Polycrystalline sample of $Nb_2PdS_5$ were prepared by solid state reaction method as described elsewhere [4]. Resistivity measurement was done using four probe method in a *Cryogenic* low temperature and high magnetic field system in conjunction with variable temperature insert. Tunnel diode oscillator technique was used to study rf penetration depth in $Nb_2PdS_5$ superconductor. This technique gives change in penetration depth with temperature [15]. It consists of a LC circuit made up of copper coil inductor and a capacitor connected in parallel and is driven to resonance using a negatively biased tunnel diode oscillator. The sample is kept inside this copper coil and cooled below its transition temperature. The frequency of the LC circuit is ~2.3MHz for our apparatus. The inductance of the coil is directly proportional to the cross sectional area occupied by the magnetic flux passing through the coil. Thus as sample temperature decreases below $T_C$, the amount of magnetic flux passing through the coil also decreases that results in a change frequency of LC circuit. This change in frequency is proportional to the magnetic susceptibility of the sample. If $f_0$ is the frequency of empty coil and $f_s$ is the frequency of coil with sample then the shift in resonant frequency $\Delta f = f_s - f_0 = -4\pi\chi G$ where $\chi$ is the total magnetic susceptibility and G is the geometric calibration defined by the geometry of coil and sample. For our apparatus we have calculated the G by taking pure niobium sample of similar dimensions as of our sample and fitting its low temperature penetration data with isotropic BCS s-wave equation. The value of G is determined to be 2.27 Å/Hz. The susceptibility in the Meissner state is related to the penetration depth with the formula as $-4\pi\chi = \frac{1}{(1-N)}\left[1 - \frac{\lambda}{R}\tanh\left(\frac{R}{\lambda}\right)\right]$ where R is the effective dimension of the sample [16].

An oven stabilized frequency counter (HP-53113A) was used for measurements of frequency, whereas, the sample temperature was measured using Lakeshore Temperature Controller 320. The ac field of the coil is ~1 μT which is much less than the lower critical field for the sample ~150 Oe thus the sample remains in Meissner state below its transition temperature.

**Result and Discussions**

Figure 1 depicts the room temperature observed and fitted X-Ray diffraction (XRD) patterns for as synthesised $Nb_2PdS_5$. The studied sample is crystallized in centro-symmetric structure with C2/m space group. The lattice parameters are a = 12.132(2) Å, b = 3.2774(3) Å, and c = 15.024(1) Å. These are in agreement with earlier reports [1-4]. The inset of Figure 1, shows the DC magnetic susceptibility of the studied $Nb_2PdS_5$ compound with superconducting transition temperature ($T_c$) around 6 K. The shielding volume fraction (ZCF) is close to 90%. It is clear that the studied $Nb_2PdS_5$ is a bulk superconducting below ~6 K. This is in agreement with earlier reports on this compound [1-4].

Figure 2 confirms superconductivity in $Nb_2PdS_5$ sample. Left axis of fig 1 shows the temperature dependence of sample frequency that shows a sharp decrease at 6.25K which is indicative of superconducting transition. The sudden decrease in frequency occurs when sample goes from superconducting to normal state thereby allowing magnetic flux to pass through it. The right axis shows electrical resistivity as a function of temperature. The resistivity shows superconducting transition with $T_{conset}$ at 6.3K and $T_{czero}$ at 5.8K.

The inset in Figure 2 shows change in penetration depth as a function of temperature normalized with the total change in penetration depth down to lowest measured temperature. For an isotropic one-gap BCS model the change in penetration depth $\Delta\lambda$ at $T \ll T_c$ follows as;

$$\Delta\lambda(T) = \lambda_0 \sqrt{\frac{\pi\Delta_0}{2k_B T}} \exp\left(-\frac{\Delta_0}{k_B T}\right) \qquad \text{Eq.1}$$

Here, $\lambda_0$ and $\Delta_0$ are the values of penetration depth $\lambda$ and the superconducting energy gap $\Delta$ at T = 0 [17,18]. The same exponential temperature dependence of $\Delta\lambda$ is valid for superconductor with a nodeless anisotropic energy gap or distinct gaps on different Fermi-surface sheets but in that case, $\lambda_0$ gives effective value that depends on the details of the gap anisotropy and $\Delta_0$ approximately equals to the minimum energy gap in the system[17]. For d-wave pairing in the clean limit,

$$\Delta\lambda(T) \approx \lambda(0)\frac{2\ln 2}{\alpha\Delta_0}T \qquad \text{Eq.2}$$

With $\alpha = \Delta_0^{-1}\left[d\Delta(\Phi)/d\Phi\right]_{\Phi \to \Phi_{node}}$ and $\Delta(\Phi)$ is the angle dependent gap function [19]. In case of dirty limit, d-wave gap is suppressed and the temperature dependence changes from linear behaviour to power law [19]; $\Delta\lambda \sim T^2$. To compare the appropriateness of these models, in Figure 3 we show low temperature change in penetration depth $\Delta\lambda$ of $Nb_2PdS_5$ along with the fittings of conventional BCS model(solid back line) and quadratic temperature dependence (dashed red line). Clearly the s-wave BCS model proves to be a better fit for the data as compared to quadratic temperature dependence. This rules out the possibility of d-wave pairing in this compound. At low temperature $T/T_c < 0.5$, fitting parameters obtained from fitting s-wave BCS model yield gap ratio $2\Delta_0/k_B T_c = 3.9$ and the corresponding value of energy gap value is $\Delta_0 \sim 1.05$meV. This value of gap ratio is slightly higher than the BCS value of 3.53 which suggests this superconductor to be an intermediate strong coupling superconductor.

The low temperature data of $\Delta\lambda$ allows us to determine the pairing symmetry and presence of nodes. But in order to detect multiple gaps or anisotropies of the superconducting gap it is important to analyse superfluid density given by $\rho(T) = [\lambda(0)/\lambda(T)]^2$ over the entire temperature range where $\lambda(T) = \Delta\lambda + \lambda(0)$. We need $\lambda(0)$ to evaluate $\lambda(T)$ and hence superfluid density but we cannot get the value of $\lambda(0)$ directly using the tunnel diode oscillator technique. Thus we use lower critical field value to estimate $\lambda(0)$ using the formula;

$$H_{c1} = \frac{\Phi_0}{4\pi\lambda^2}\left(\ln\frac{\lambda}{\xi} + 0.485\right) \qquad \text{Eq. 3}$$

where the value of $H_{c1}(0)$ and $\xi(0)$ are given by ~ 150 Oe and 37 Å respectively [2,20], and $\Phi_0$ is magnetic flux quantum = $2 \times 10^{-7}$ Gcm$^2$. This gives $\lambda(0) = 220$nm.

The normalized value of superfluid density $\rho(T) = [\lambda(0)/\lambda(T)]^2$ takes the following form in dirty and clean limit;

$$\left.\frac{\lambda^2(0)}{\lambda^2(T)}\right|_{dirty} = \frac{\Delta(T)}{\Delta_0}\tanh\left[\frac{\Delta(T)}{2k_BT}\right] \qquad \text{Eq. 4}$$

$$\left.\frac{\lambda^2(0)}{\lambda^2(T)}\right|_{clean} = 1 + 2\int_{\Delta(T)}^{\infty}\left[\frac{\partial f}{\partial E}\right]\frac{E}{\sqrt{E^2 - \Delta(T)^2}}dE \qquad \text{Eq. 5}$$

Here $f = [1+\exp(E/k_BT)]^{-1}$ is the Fermi function. The temperature dependence of gap function $\Delta(T)$ calculated from the simple, weak-coupling, isotropic, BCS model gives

$$\Delta(T) = \Delta_0\tanh\left[\frac{\pi k_BT_c}{\Delta_0}\sqrt{a\left(\frac{T_c}{T}-1\right)}\right] \qquad \text{Eq.6}$$

Here $\Delta_0$ is the gap magnitude at zero temperature and '*a*' is a free parameter that depends on the particular pairing state and $k_B$ is Boltzmann constant [17]. The integration is over the entire quasi-particle energies measured from the chemical potential. Figure 5 shows the superfluid density over the entire temperature range using $\lambda(0) = 220$ nm and a comparison with two fluid Gorter and Casimir model and the clean and dirty s-wave models. The data matches very well to the dirty s-wave fitting. Taking '$\Delta_0$' and '*a*' as a free parameter, we have fitted the data with Eq. 5 using *Mathematica* software which gives the value of Superconducting gap ratio $2\Delta_0/k_B T_c = 4 \pm 0.127$ and the corresponding value of energy gap value $\Delta_0 \sim 1.07$ meV and $a = 1.1 \pm 0.014$. We have also tried to fit *alfa* model in our sample for two gap superconductivity which was earlier used in $MgB_2$ to give two gaps. The result from this model converges to single gap equation which suggests single gap in this compound. The recent report of μSR study on this compound also confirms nodeless s-wave superconductivity [21].

**Conclusion**

We have studied penetration depth measurement in $Nb_2PdS_5$ using tunnel diode oscillator technique. The low temperature penetration depth data is best described by BCS s-wave model with gap ratio $2\Delta_0/k_B T_c = 3.9$ and the corresponding value of energy gap value $\Delta_0 \sim 1.05$ meV. Super fluid density measurements over the entire temperature range indicate single gap superconductivity in this compound. In essence our results rule out unconventional superconductivity in the record high upper critical field superconductor $Nb_2PdS_5$.

**Figure Captions**

Figure 1 Room temperature XRD pattern of $Nb_2PdS_5$, inset shows both Zero-field cooled (ZFC) and Field cooled (FC) magnetic susceptibility.

Figure 2 Left axis shows the temperature dependence of frequency shift with superconducting transition at 6.25 K. Right axis shows the temperature dependence of resistivity with $T_{c\,onset}$ at 6.3 K and $T_{c\,zero}$ at 5.8 K. Inset shows change of penetration depth $\Delta\lambda$ normalized to total shift down to 1.68K that reveals a sharp superconducting transition at $T_c \sim 6.25$ K .

Figure 3 Low temperature penetration depth $\Delta\lambda$ dependence of $Nb_2PdS_5$ as a function of reduced temperature. Solid black line shows the exponential fitting using Eq. 1 with fitting parameter as $2\Delta_0/k_BT_c = 3.9 \pm 0.18$ and dotted red line is the $T^2$ fitting of the data.

Figure 4 Plot showing a comparison of temperature dependence of superfluid density of $Nb_2PdS_5$ with clean s-wave, dirty s-wave and two fluid models over the entire temperature range. Red line is the best fitting for a single s-wave equation using Eq. 5 with fitting parameter $2\Delta_0/k_BT_c = 4 \pm 0.128$ and $a = 1.1 \pm 0.014$.

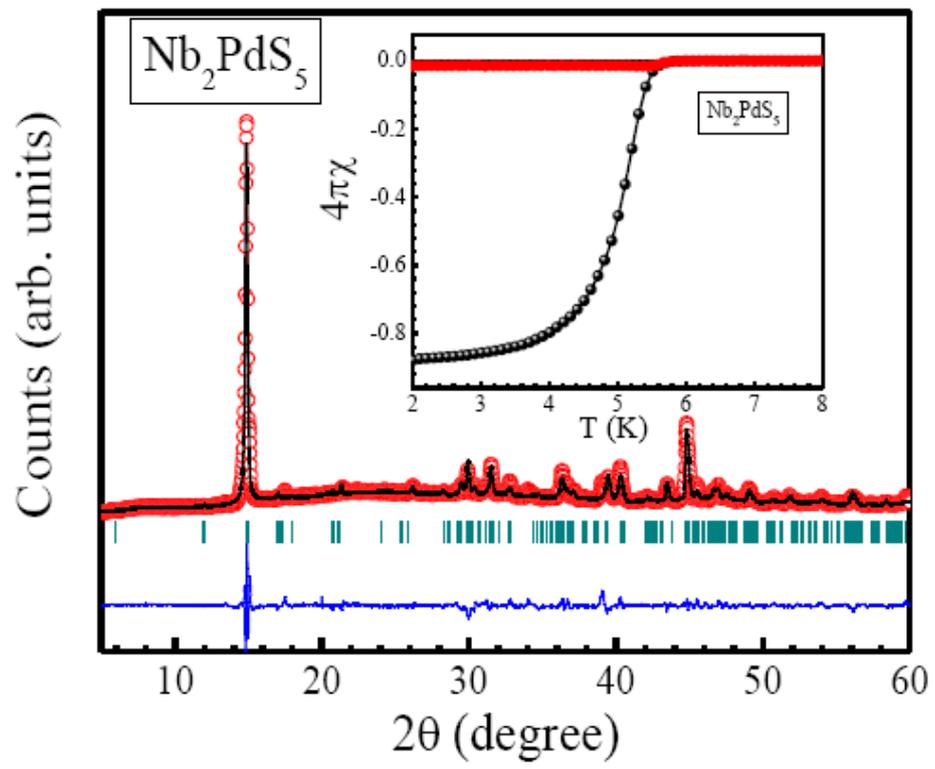

**Figure 1**

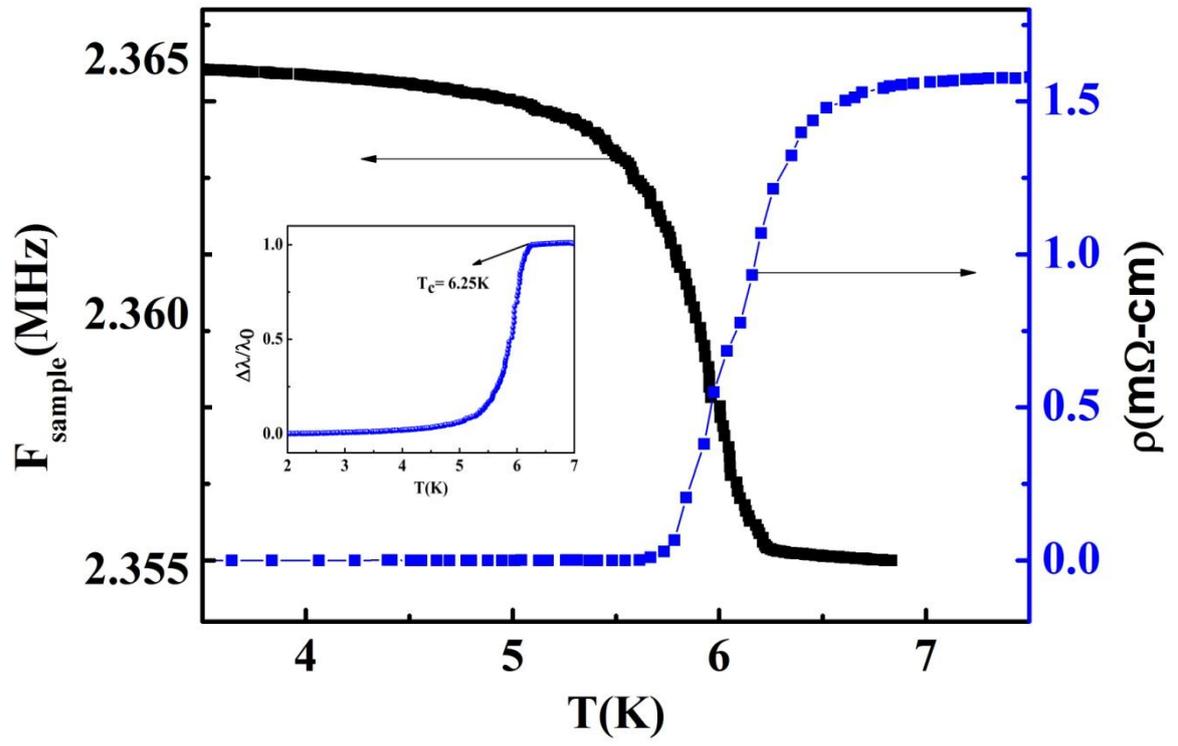

**Figure 2**

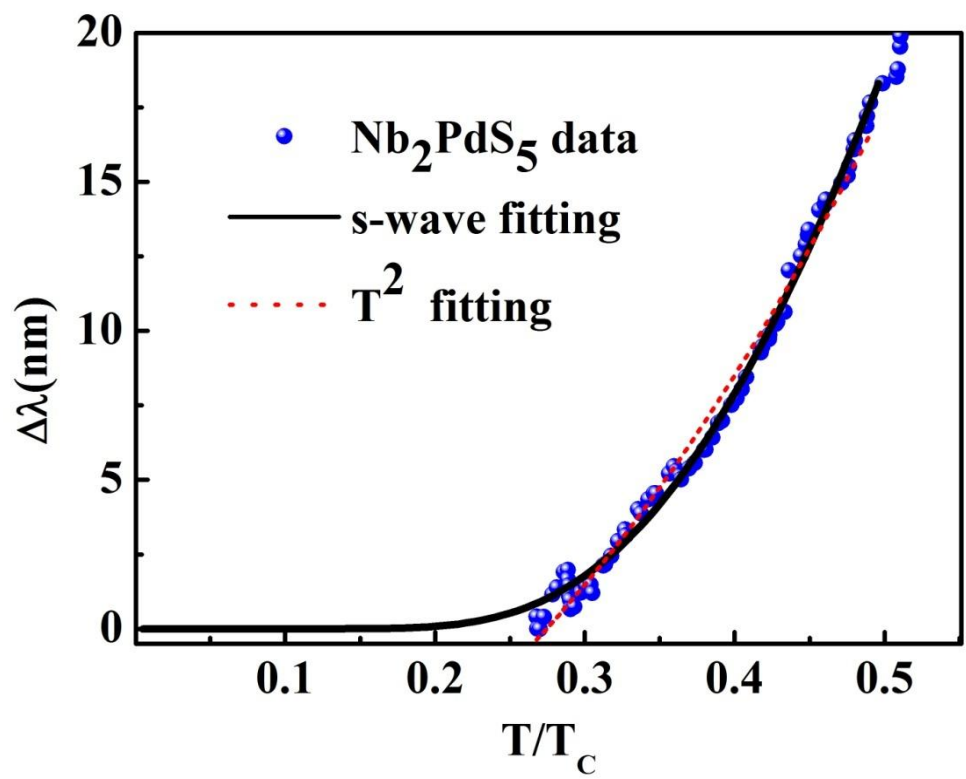

**Figure 3**

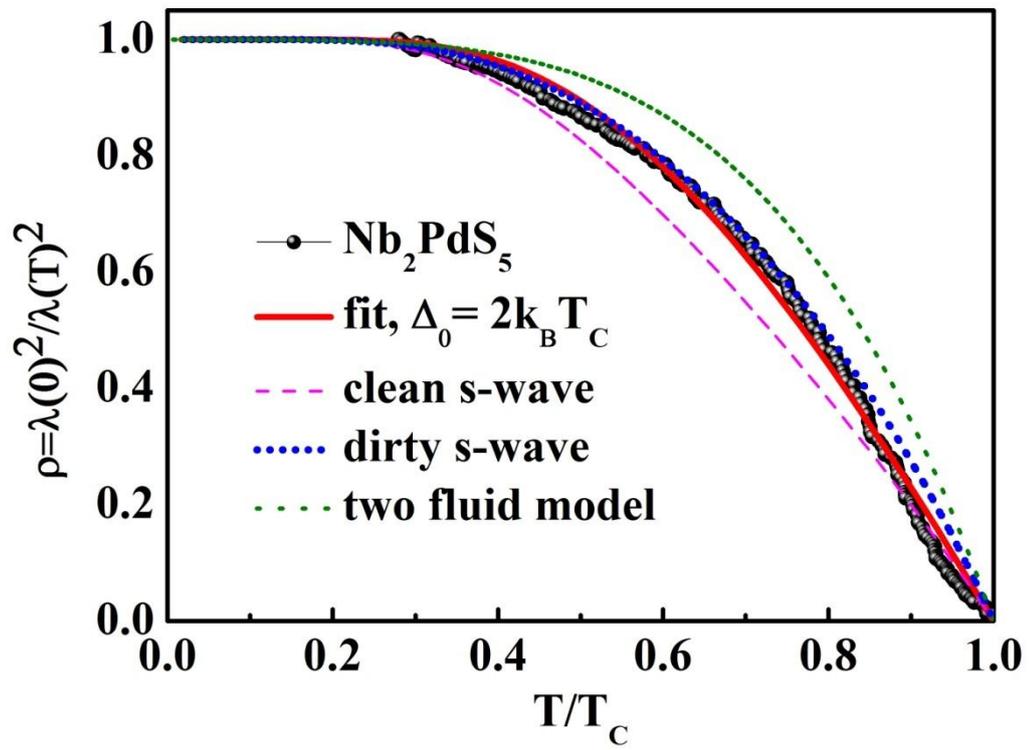

**Figure 4**